\documentclass[referee]{raa}           
\usepackage{float}
\usepackage{graphicx,times}
\usepackage{natbib}
\usepackage{amssymb,amsmath}
\bibpunct{(}{)}{;}{a}{}{,}

\usepackage[pagebackref=true]{hyperref}

\begin{document}

   \title{Burst phase distribution of SGR J1935+2154 based on \textit{Insight}-HXMT}

 \volnopage{ {\bf 20XX} Vol.\ {\bf X} No. {\bf XX}, 000--000}
   \setcounter{page}{1}

   \author{Xue-Feng Lu
   \inst{1,2}, Li-Ming Song\inst{1,2}, Ming-Yu Ge\inst{1}, 
      You-Li Tuo\inst{1}, Shuang-Nan Zhang\inst{1,2}, Jin-Lu Qu\inst{1}, Ce Cai\inst{3}, Sheng-Lun Xie\inst{1}, Cong-Zhan Liu\inst{1}, Cheng-Kui Li\inst{1}, Yu-Cong Fu\inst{1}, Ying-Chen Xu\inst{1,2}, Tian-Ming Li\inst{1,2}  
      }

   \institute{ Key Laboratory of Particle Astrophysics, Institute of High Energy Physics, Beijing 100049, 
China; {\it luxf@ihep.ac.cn}\\
	\and
     University of Chinese Academy of Sciences, Chinese Academy of Sciences, Beijing 100049, China\\
\and
College of Physics and Hebei Key Laboratory of Photophysics Research and Application,
Hebei Normal University, Shijiazhuang, Hebei 050024, China\\
\vs \no
   {\small Received 20XX Month Day; accepted 20XX Month Day}
}

\abstract{On April 27, 2020, the soft gamma ray repeater SGR J1935+2154 entered its intense outburst episode again. \textit{Insight}-HXMT carried out about one month observation of the source. A total number of 75 bursts were detected during this activity episode by \textit{Insight}-HXMT, and persistent emission data were also accumulated. We report on the spin period search result and the phase distribution of burst start times and burst photon arrival times of the \textit{Insight}-HXMT high energy detectors and Fermi/Gamma-ray Burst Monitor (GBM). We find that the distribution of burst start times is uniform within its spin phase for both \textit{Insight}-HXMT and Fermi/GBM observations, whereas the phase distribution of burst photons is related to the type of a burst's energy spectrum. The bursts with the same spectrum have different distribution characteristics in the initial and decay episodes for the activity of magnetar SGR J1935+2154. 
}

   \authorrunning{X.-F. Lu et al. }            
   \titlerunning{Burst phase distribution of SGR J1935+2154}  
   \maketitle

%
\section{Introduction}           
\label{sect:intro}
Magnetars are a class of special celestial objects with super strong magnetic field (often more than $10^{14}$ G) in the universe, which are usually thought of as young neutron stars. Compared to conventional pulsars, magnetars are characterized by intense energetic phenomena in X-ray band and soft gamma ray band (\citealt{Kaspi+etal+2017}). This is why magnetars are commonly classified as anomalous X-ray pulsars (AXPs) and soft gamma repeaters (SGRs) (\citealt{Scholz+Kaspi+2011,Woods+etal+2005}). It is widely believed that magnetars are powered by the decay of their supercritical magnetic fields, perhaps an external magnetic field (\citealt{Kouveliotou+etal+1998}) or an internal magnetic field (\citealt{Thompson+Duncan+1995}).

SGR J1935+2154 was discovered in 2014 when Swift-BAT (Burst Alert Telescope) was triggered by short bursts from Galactic plane (\citealt{Stamatikos+etal+2014}). Subsequent Chandra observations located the burst from the direction of the supernova remnant G57.2+0.8 (\citealt{Kothes+etal+2018}). Based on Chandra and XMM-Newton data, a spin period of 3.24 s and spin-down rate of $1.43(1) \times 10^{-11}\ \rm{s\ s^{-1}}$ was discovered, which implying a surface bipolar magnetic field strength of approximately $2.2 \times 10^{14}$ G (\citealt{Israel+etal+2016}). Combined with its burst characteristics, the source was identified as a magnetar. In 2015, 2016 and 2019, SGR J1935+2154 have many burst activity episodes, releasing a lot of energy in persistent and burst emission (\citealt{Younes+etal+2017,Lin+etal+2020a}).

Since 2020 April 27, SGR J1935+2154 entered into its active episode again, and multiple X-ray and gamma-ray telescopes detected a large number of intense bursts. Several hours after the outburst onset, CHIME known as Canadian Hydrogen Intensity Mapping Experiment (\citealt{CHIME_Collaboration+etal+2020}) and STARE2 known as The Survey for Transient Astronomical Radio Emission 2 (\citealt{Bochenek+etal+2020}) detected an intense fast radio burst (FRB) from the source direction, respectively. At the same time, multiple hard X-ray telescopes (\citealt{Mereghetti+etal+2020,Li+etal+2021,Ridnaia+etal+2021,Younes+etal+2021}) detected the hard X-ray signal from the source that was associated with this FRB. The correlation between X-ray burst and FRB provides evidence that at least some FRBs can be originated from magnetars. After initial 2020 active episode, NICER, Fermi and \textit{Insight}-HXMT performed long-term observations of the source evolution. (\citealt{Younes+etal+2020}) presented observations of a burst storm and long-term persistent emission evolution of the SGR J1935+2154 based on NICER data. They find a double-peaked pulse profile of soft X-ray emission of the source, corresponding to a frequency $f=0.307946(2)\ \rm{Hz}$. The burst peak arrival times detected by NICER in 1 keV to  10 keV follow a uniform distribution in pulse profile. (\citealt{Kaneko+etal+2021}) presented the results of time-resolved spectral analysis of the ``burst forest" lasted for 130 s observed by Fermi/Gamma-ray Burst Monitor. They converted the GBM photon arrival times to barycentric times and studied the lightcurve and the spectral parameter evolution with the NICER pulse profile. The results show that the Comptonized model (COMPT, a power law with exponentially cutoff) fits these bursts with an anti-correlation, i.e., the spectra with high peak energy $E_{\rm peak}$ appear at or close to the minima of the pulse profile. They also noted that even though the flux varied by two orders of magnitude the single blackbody kT remains constant around 7 keV and the double blackbodies high kT also remains roughly constant at about 14 keV. In (\citealt{Lin+etal+2020b}), temporal and time-integrated analysis of the 125 bursts (excluding the 130 s burst forest) of SGR J1935+2154 2020 actvie episode detected by Fermi/GBM was reported. They found a growing trend for the evolution of the total burst fluence since its discovery in all active episodes. They also studied the last time evolution of the burst, finding a similar log-Gaussian distribution as other magnetars.

On April 28, approximately 13 hours after the outburst episode onset, \textit{Insight}-HXMT started observation of this source. This observation lasted 33 days, and a total number of 75 bursts were detected. In (\citealt{Cai+etal+2022b}), they reported similar results as (\citealt{Lin+etal+2020b}) for these 75 bursts, which are on average much fainter than the GBM bursts reported in (\citealt{Lin+etal+2020b}). In this letter, we study on the burst phase evolution properties of Fermi/GBM 125 bursts and \textit{Insight}-HXMT 75 bursts, and also the 2020 persistent emission based on \textit{Insight}-HXMT data. Section \ref{sect:Obs} gives the HXMT data reduction methods, Section \ref{sect:Result} gives the analysis results, and Section \ref{sect:discussion} presents the discussion.


\section{Observaion and data reduction}
\label{sect:Obs}
\textit{Insight}-HXMT was launched on 2017 June 15, which carries three collimated telescopes covering 1–10 keV (the Low Energy X-ray telescope, LE, geometrical area of 384 cm$^{2}$), 5–30 keV (the Medium Energy X-ray telescope, ME, geometrical area of 952 cm$^{2}$) and 20–250 keV (the High Energy X-ray telescope, HE, geometrical area of about 5000 cm$^{2}$) (\citealt{Zhang+etal+2020,Liu+etal+2020,Chen+etal+2020,Cao+etal+2020}). The Target of Opportunity (ToO) observation for \textit{Insight}-HXMT of SGR J1935+2154 lasted from April 28 07:14:51 UTC to June 1 00:00:01 UTC with a total effective exposure of 1650 ks. The detailed observation time list can be found in (\citealt{Cai+etal+2022a}). We refined the \textit{Insight}-HXMT burst information with the same data reduction method as shown in (\citealt{Cai+etal+2022a}). For Fermi/GBM we used the same data reduction and burst refining method as shown in (\citealt{Lin+etal+2020b}). For phase analysis, all the burst data have been converted to barycentric times.

The event data are processed with the \textit{Insight}-HXMT Data Analysis Software package (HXMTDAS) version 2.05. Standard data processing is used for HE, ME and LE event data. First \textit{hepical}, \textit{mepical} and \textit{lepical} are used to calibrate event photons of HE, ME and LE, respectively, with Calibration Database (CALDB) of \textit{Insight}-HXMT. Then good time interval is done directly for HE calibrated data with \textit{hegtigen}. For LE data with the two-split events reconstruction and classification are executed with \textit{lerecon} first and then \textit{legtigen}. For ME data, \textit{megrade} is used to calculate event grade and dead time correction before \textit{megtigen}. Finally, \textit{hescreen}, \textit{mecreen} and \textit{lescreen} are used to do good time data extraction and \textit{hxbary} is used to do solar system centroid correction for subsequent phase analysis. Then all the burst signals and spurious pulse signals are thoroughly removed to do spin period search. Based on NuSTAR (\citealt{Borghese+etal+2020}) and NICER (\citealt{Younes+etal+2020}) results, SGR J1935+2154 persistent emission mainly concentrated in the low energy range, thus the X-ray pulse profile analysis of the source only use the \textit{Insight}-HXMT LE data. 

\section{Result}
\label{sect:Result}
\subsection{X-ray pulse profile analysis based on \textit{Insight}-HXMT LE data}
\begin{figure}
   \centering
   \includegraphics[width=8cm, angle=0]{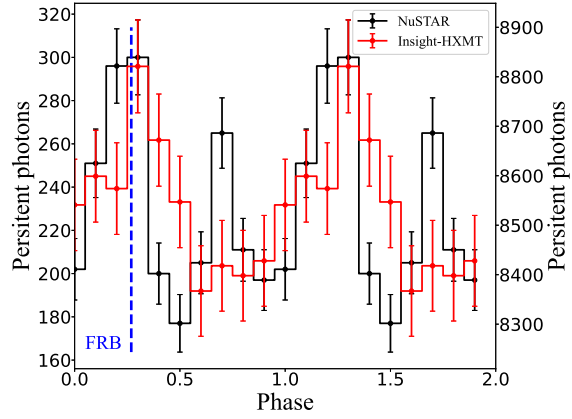}
   \caption{Comparing the persistent X-ray pulse profile of \textit{Insight}-HXMT with NuSTAR, using the search period $f=0.3079452\ \rm{Hz}$ ($T_{0}=58967.423047\ \rm{MJD}$) of NuSTAR on May 2, the data of \textit{Insight}-HXMT is used from May 1 to May 5 in the energy range 1.5 keV to 6.8 keV. The black line is the NuSTAR 1.0 keV to 6.0 keV result, and the red line is the \textit{Insight}-HXMT result. The blue dashed vertical line represent the phase of FRB 200428. The left axis is NuSTAR scale, and the right axis is the \textit{Insight}-HXMT scale.}
   \label{Fig1}
   \end{figure}

To check \textit{Insight}-HXMT detection ability for the persistent emission of SGR J1935+2154, we performed a preliminary examination of the low-energy data from 1.5 keV to 6.8 keV of \textit{Insight}-HXMT based on the periodic search results $f=0.3079452\ \rm{Hz}$ ($T_{0}=58967.423047 \ \rm{MJD}$) on 2020 May 2 of NuSTAR observations reported in (\citealt{Borghese+etal+2020}). With the \textit{Insight}-HXMT LE data from May 1 to May 5, we folded the pulse profile and found that \textit{Insight}-HXMT could give similar pulse profiles under the periodic parameters given by NuSTAR, as shown in Figure \ref{Fig1}, where the red line represents the \textit{Insight}-HXMT result, whereas the black line represents the NuSTAR result. FRB 200428 phase is also marked in the figure with the blue dashed line. We can see that \textit{Insight}-HXMT does detect persistent radiation photons from the magnetar SGR J1935+2154 during its observations. We also find that the pulse profile observed by \textit{Insight}-HXMT shows a little deviation compared to NuSTAR, which may be due to the long time span of the \textit{Insight}-HXMT data and possible evolution of the period, and mainly due to the low signal to noise ratio of \textit{Insight}-HXMT LE data, as shown in the figure right scale, \textit{Insight}-HXMT has a very high background and this will cause large fluctuations in the profile.

Subsequently, we carried out a detailed study on the one month observation data of \textit{Insight}-HXMT, and searched the spin period of SGR J1935+2154 based on \textit{Insight}-HXMT LE data. Considering the rapid evolution of magnetar spin period and the fluence decay of the persistent radiation after the burst episode onset, we grouped the data of \textit{Insight}-HXMT. Finally, we found the periodic signal in the data segment MJD 58969.29054398 to MJD 58972.22856481. We restricted our search interval to the frequency range $0.3079\ \rm{Hz}< f <0.3080\ \rm{Hz}$, which covers the source spin frequency $f=0.3079462\ \rm{Hz}$ reported in (\citealt{Younes+etal+2020}) with NICER data. We found the largest $\chi^{2}$ value of 33.2 (corresponding to the largest $Z_1^2=30.2$, $Z^2$ is conceptually similar to the $\chi^{2}$ but has high values when the signal is well described by a small number of sinusoidal harmonics, the specific expression can be found in (\citealt{Buccheri+etal+1983})) at frequency $f=0.3079433(14)\ \rm{Hz}$, which corresponds to an spin period of $3.24735(2)\ \rm{s}$.

Due to the large margin of error, here we just give a simple result of \textit{Insight}-HXMT pulse period search. Also because the long time span of our study, none of the existing ephemeris can fully cover all the bursts. The ephemeris used in this paper are fitted results based on the combined ephemeris evolution data derived from NuSTAR, NICER and XMM-Newton in (\citealt{Ge+etal+2022}), which yield $f=0.30794447(1)\ \rm{Hz}$, $\dot{f} = -2.165 \times 10^{-12}\ \rm{Hz\ s^{-1}}$ at an epoch $T_{0}=58967\ \rm{MJD}$; this ephemeris can cover the whole observation period of \textit{Insight}-HXMT, and is used in the subsequent study of the burst phase distribution for the bursts detected with Fermi/GBM and \textit{Insight}-HXMT.

\subsection{Burst phase characteristics}
\begin{figure}
   \centering
   \includegraphics[width=8cm, angle=0]{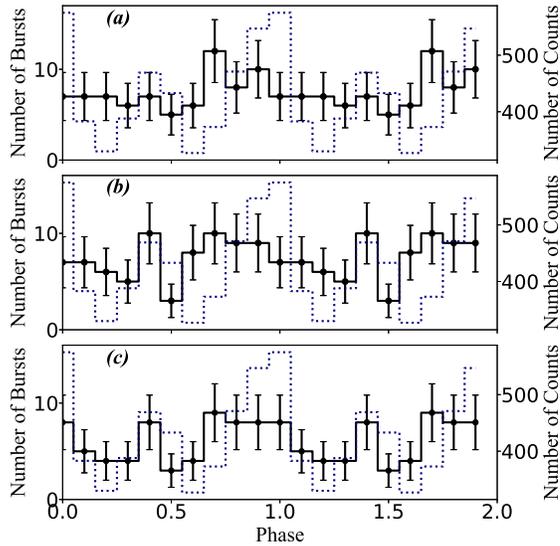}
   \caption{The phase distribution of the burst start times. Panel (a) is the HXMT HE telescope data, panel (b) is the HXMT ME telescope data, panel (c) is the HXMT LE data. In each panel, the black solid curve is HXMT data, the dotted light blue curve is the X-ray pulse profile conducted with NuSTAR May 2, 2020 data.}
   \label{Fig2}
   \end{figure}

During \textit{Insight}-HXMT 33 days observation of SGR J1935+2154, only 61 bursts are detected by LE telescope, and most of them are very weak with only several photons. But for HE telescope there are 75 remarkable bursts, while for ME telescope there are 74 bursts. Because HE has more complete samples, and LE and ME burst photons have consistent statistical characteristics as HE, we only use HE bursts data in the subsequent analysis. Since the energy spectra and flux information about the bursts have been reported in (\citealt{Cai+etal+2022b}), this work mainly analyzes the phase distribution characteristics of these bursts. In the paper, the burst start time is defined as the start time of the first Bayesian block (\citealt{Scargle+etal+2013}) within the burst time window. First, a statistical analysis of the start time for each \textit{Insight}-HXMT telescope burst is performed, as shown in Figure \ref{Fig2}, where panel (a) shows HE bursts, panel (b) shows ME bursts and panel (c) shows LE bursts, the burst start times for the three telescopes of \textit{Insight-HXMT} follow the same distribution; the $\chi^{2}$ test is used to check any structure significance, but no significant difference is found from a uniform distribution (with equivalent Gaussian significance less than $1\sigma$).

\begin{figure}
   \centering
   \includegraphics[width=8cm, angle=0]{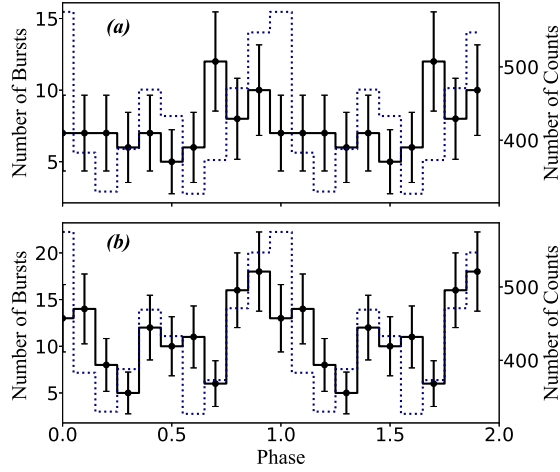}
   \caption{The distribution of burst start time in phase space for Fermi/GBM and \textit{Insight}-HXMT. Panel (a) is for all the 75 bursts start time of \textit{Insight}-HXMT, panel (b) is for the 113 bursts start time of Fermi. The solid black line is the burst numbers, and the dotted light blue line is the X-ray pulse profile with the NuSTAR May 2 data. }
   \label{Fig3}
   \end{figure}

According to (\citealt{Lin+etal+2020b}) and our search result, during the 2020 April activity episode of SGR J1935+2154, Fermi/Gamma-ray Burst Monitor found a total number of 125 bursts (from April 27 to May 20, excluding the 130 s ``burst forest"). The 12 bursts from April 28 06:00:00 UTC to May 20 are excluded from these 125 bursts, and the rest 113 bursts are used to compare with \textit{Insight}-HXMT observations; the Fermi/GBM burst episode here is deemed as an intense activity stage. For comparison, the bursts period of \textit{Insight}-HXMT observation is called the decay activity stage. Thus we have a complete separation of Fermi/GBM and \textit{Insight}-HXMT bursts in time. The start time of each burst and the photon arrival times within the burst are converted to barycentric time and then phase. The burst start time distribution in phase for the two stages is shown in Figure \ref{Fig3}. The top panel shows the start time distribution of the 75 bursts detected with \textit{Insight}-HXMT HE, and the bottom panel shows the start time distribution of the 113 bursts detected with Fermi/GBM. The dotted light blue line is the X-ray pulse profile based on NuSTAR May 2 data. After $\chi^{2}$ test, neither Fermi/GBM nor \textit{Insight}-HXMT detected any  significant structure in phase distribution (with equivalent Gaussian significance less than $2\sigma$).
  
\begin{figure}
   \centering
   \includegraphics[width=8cm, angle=0]{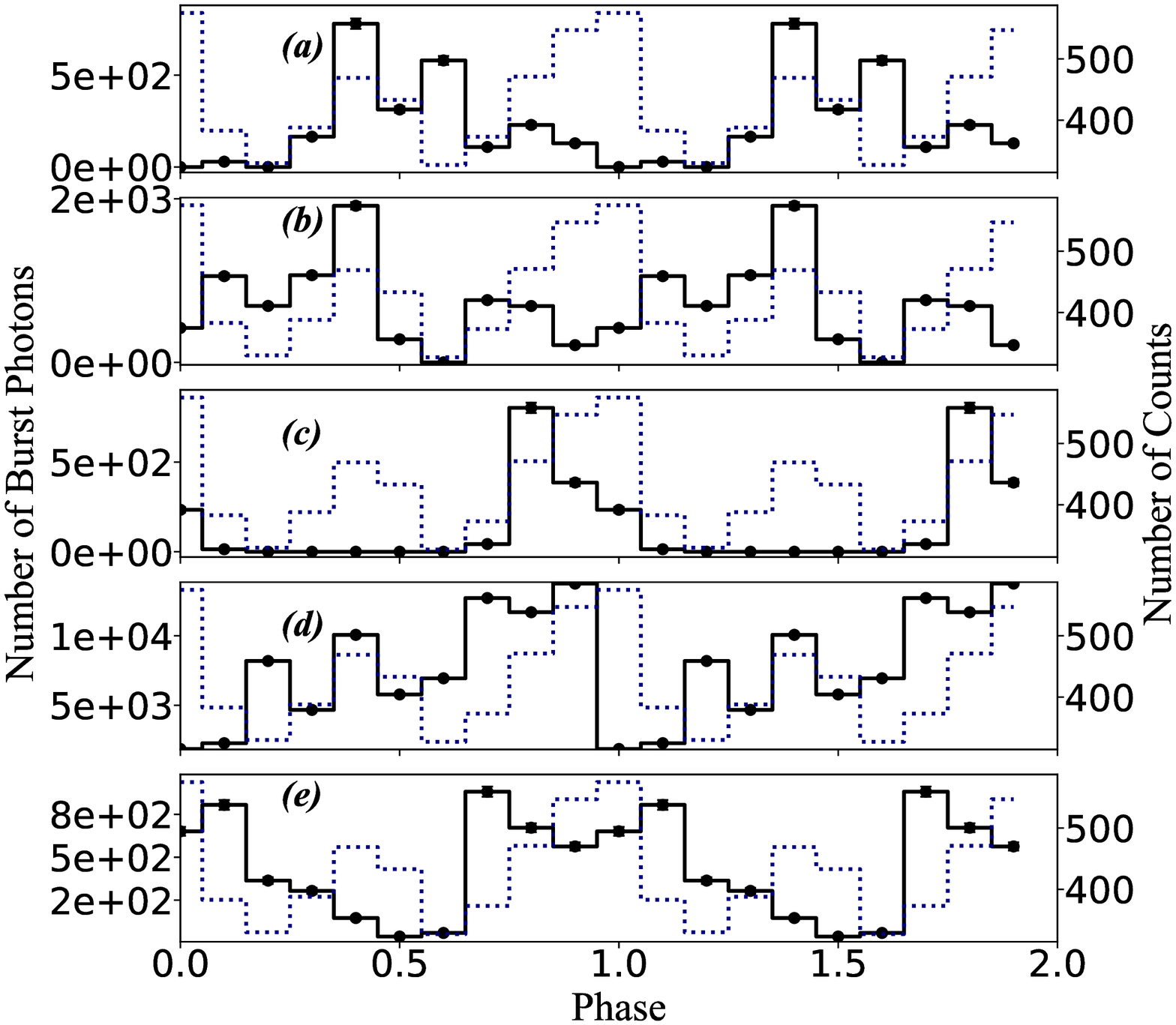}
   \caption{The phase distribution of \textit{Insight}-HXMT burst photons for different spectra models. Panel (a) is for bursts with BB spectra, panel (b) is for bursts with BB+BB spectra, panel (c) is for bursts with BB+PL spectra, panel (d) is for bursts with CPL spectra, panel (e) is for bursts with PL spectra. For each panel, the black solid line represents burst photon numbers, while the dotted lightblue line is the X-ray pulse profile of SGR J1935+2154 based on NuSTAR May 2 observation.}
   \label{Fig4}
   \end{figure}

 The energy spectrum of magnetar bursts is complex and diverse, and different types of bursts may show unique distribution characteristics in phase due to different generating mechanisms. To study the phase distribution characteristics of burst photons with different types of energy spectrum, we classify the bursts according to the energy spectrum fitting model actually used by \textit{Insight}-HXMT and Fermi/GBM. In the energy spectrum fitting analysis of the \textit{Insight}-HXMT's 75 bursts, the cut off power law (CPL), double blackbodies (BB+BB), blackbody plus power law (BB+PL), single blackbody (BB) and power law (PL) fitting models are used; the phase distribution of burst photon arrival time for each type model is investigated respectively. The result is shown in Figure \ref{Fig4}, panel (a) shows bursts with BB spectra, panel (b) shows bursts with BB+BB spectra, panel (c) shows bursts with BB+PL spectra, panel (d) shows bursts with CPL spectra, and panel (e) shows bursts with PL spectra. For each panel, the black solid line represents burst photon number, and the dotted light blue line is the X-ray pulse profile of SGR J1935+2154 based on NuSTAR May 2 observation. We find that the phase distribution of burst photons with power law spectrum component showed a tendency to align with the main peak of the persistent radiation profile, especially the single PL spectrum feature is the most obvious, almost all the burst photons are concentrated to the main peak. For bursts with BB spectrum their burst photons tend to synchronize with the secondary peak, especially the single BB bursts. For BB+PL burst photons, their characteristics are dominated by the single PL feature, which means the phase distribution of burst photon arrival time is synchronized with the main persistent peak.
   
   \begin{figure}
   \centering
   \includegraphics[width=8cm, angle=0]{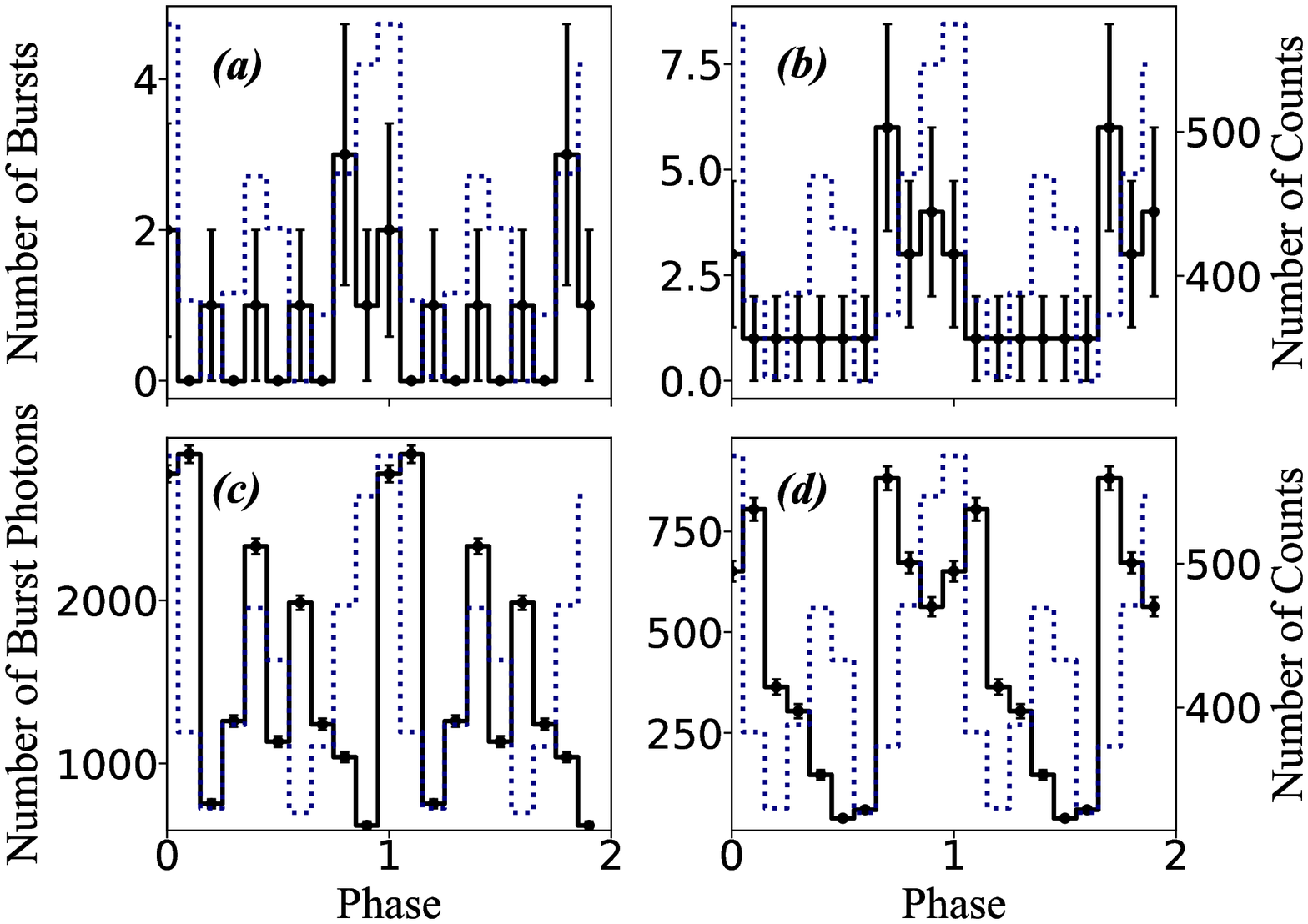}
   \caption{The burst start time and photon arrival time distribution in phase of Fermi/GBM and \textit{Insight}-HXMT power law bursts. Left: Fermi/GBM 20-200 keV data, right: \textit{Insight}-HXMT 20-200 keV data. Panels (a) and (b) are burst start time, the solid black curve is for burst number; (c) and (d) are for burst photon arrival time, where the solid black curve is for burst photon number. In each panel the dotted light blue curve  is the X-ray pulse profile based on NuSTAR May 2 observation. All the time have been converted to barycentric time and then phase.}
   \label{Fig5}
   \end{figure}
 
\begin{figure}
   \centering
   \includegraphics[width=8cm, angle=0]{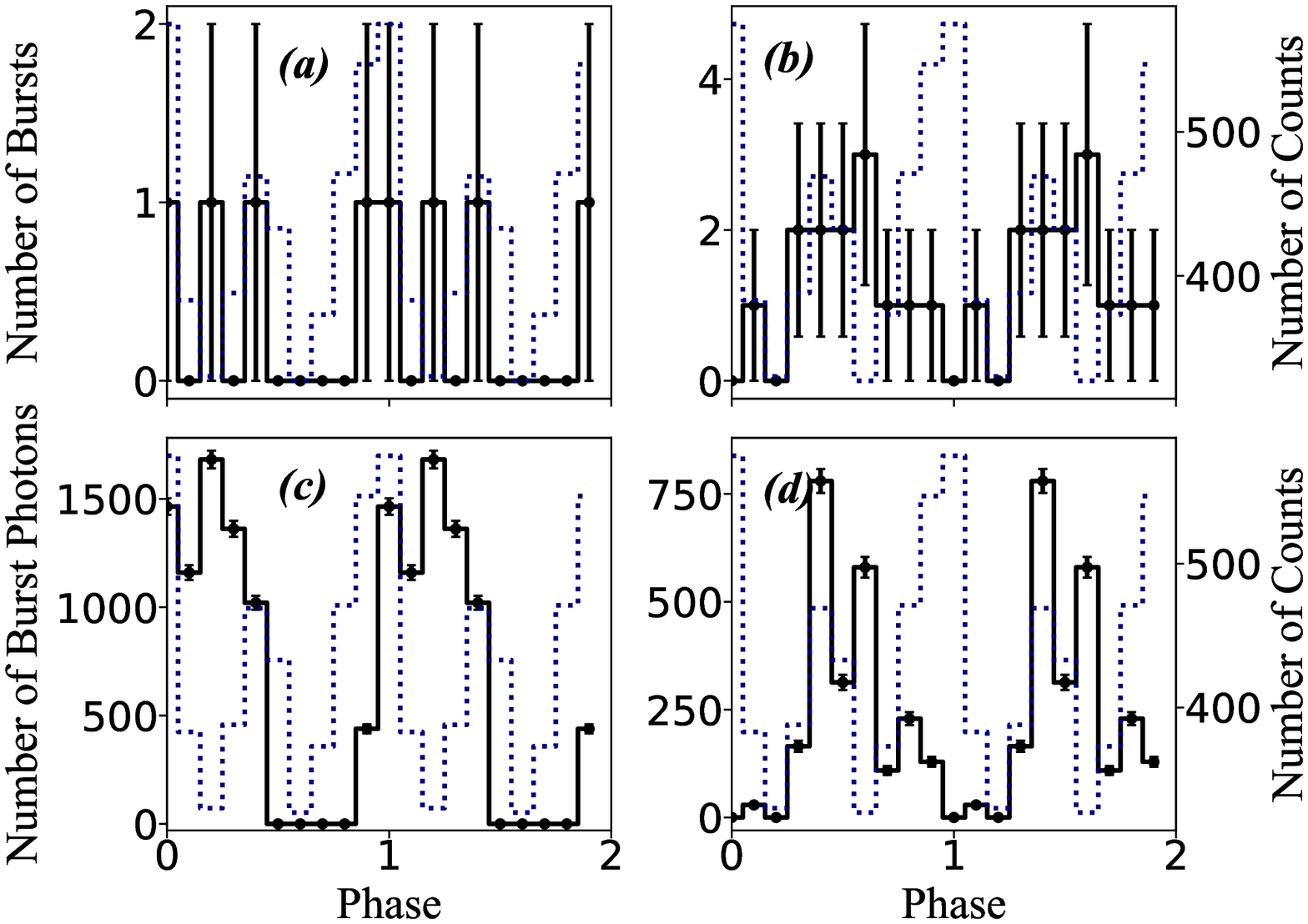}
   \caption{The burst start time and photon arrival time distribution in phase of Fermi/GBM and \textit{Insight}-HXMT for BB bursts. Left: Fermi/GBM 20-200 keV data, right: \textit{Insight}-HXMT 20-200 keV data. Panels (a) and (b) are burst start time; the solid black curve is for burst numbers; (c) and (d) are burst photon arrival time, where the solid black curve is for burst photon numbers. In each panel the dotted light blue curve  is the X-ray pulse profile based on NuSTAR May 2 observation. All the time have been converted to barycentric time and then phase.}
   \label{Fig6}
   \end{figure}

\begin{figure}
   \centering
   \includegraphics[width=8cm, angle=0]{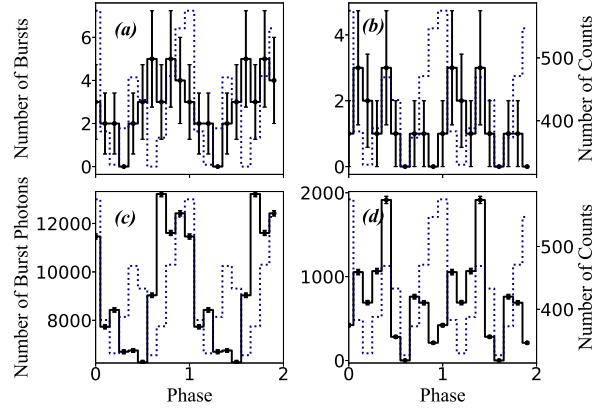}
   \caption{The burst start time and photon arrival time distribution in phase of Fermi/GBM and \textit{Insight}-HXMT for BB+BB bursts. Left: Fermi/GBM 20-200 keV data, right: \textit{Insight}-HXMT 20-200 keV data. Panels (a) and (b) are burst start time, the solid black curve is for burst number; (c) and (d) are burst photon arrival time, where the solid black curve is for burst photon number. In each panel the dotted light blue curve  is the X-ray pulse profile based on NuSTAR May 2 observation. All the time have been converted to barycentric time and then phase.}
   \label{Fig7}
   \end{figure}

 In (\citealt{Lin+etal+2020b}), the bursts spectra observed by Fermi/GBM have fitting models of power law, cut off power law, optically thin thermal bremsstrahlung (OTTB), single black body and double black body. Some bursts can be clarified by the spectrum type, and there are also a large amount of bursts which can be fitted either by CPL or BB+BB.  The numbers of each type of bursts for the \textit{Insight}-HXMT and Fermi-GBM are shown in Table \ref{tab1}. Here we choose the common model PL, BB and BB+BB to compare for the \textit{Insight}-HXMT and Fermi/GBM burst properties. As we can see in Figures \ref{Fig5} to \ref{Fig7}, for both PL and BB bursts, Fermi/GBM observed a relatively small number, and there is no significant structure in the distribution of the burst start times of these two types. But for burst photon arrival time, Fermi/GBM BB burst photons are obviously concentrated in the transition valley from the main peak to the secondary peak, while for \textit{Insight}-HXMT bursts there are almost no photons in the same phase. Fermi/GBM PL burst photons are basically synchronized with the persistent phase profile, showing a double-peak structure, the same PL bursts for \textit{Insight}-HXMT shown a similar distribution with all the photons concentrated to the same position of the persistent main peak. The burst photons of Fermi/GBM BB+BB burst distribute in the same position as the main peak, whereas \textit{Insight}-HXMT bursts have few photons in the same position. No significant distribution characteristics are found for the burst start time of either spectrum type burst.
\begin{table}
\bc
\begin{minipage}[]{100mm}
\caption[]{Number of bursts of different spectrum type\label{tab1}}\end{minipage}
\setlength{\tabcolsep}{1pt}
\small
 \begin{tabular}{c|c|c|c|c|c|c|c|cc}
  \hline
Telescope&BB& BB+BB& BB+PL& PL& CPL& CPL$\backslash$BB+BB$^{a}$& OTTB$^{b}$& Total number\\
  \hline
  \textit{Insight}-HXMT& 13& 13& 3& 22& 24& 0 & 0 &75\\
  \hline
  Fermi/GBM& 4& 29& 0& 9& 4& 39& 26&113\\
  \hline
\end{tabular}
\ec
\tablecomments{0.86\textwidth}{$^{a}$ For Fermi/GBM there are 39 bursts whose spectra can be perfectly fitted with both CPL and BB+BB.\\
     $^b$ OTTB is optically thin thermal bremsstrahlung.\\}
\end{table}

 In summary, at the intense activity stage of the magnetar SGR J1935+2154 outburst (shown by Fermi/GBM observation), there is an obvious double-peak structure in the phase of PL burst photons. As the magnetar enters the activity decay stage (shown by \textit{Insight}-HXMT observation), almost all of the power law photons concentrate near the main peak. However, the black-body and double-black-body burst photons have obvious concentrated distribution at the beginning of the magnetar activity, yet in the decay episode of magnetar activity, the photon distribution characteristics are weakened, but there is still an obvious change in phase compared with the initial stage.
 \subsection{Hardness ratio of burst photons}
\begin{figure}
   \centering
   \includegraphics[width=8cm, angle=0]{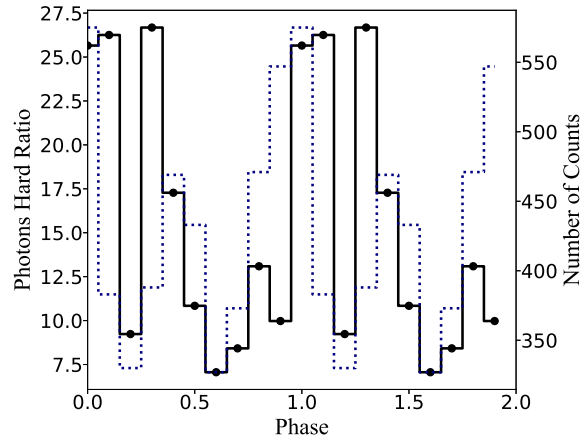}
   \caption{The hardness ratio of the high energy 30-250 keV to the medium energy 10-30 keV of the \textit{Insight}-HXMT bursts. The solid black curve is for the hardness ratio, where the dotted light blue curve is for SGR J1935+2154 X-ray pulse profile conducted from NuSTAR May 2 data.}
   \label{Fig8}
   \end{figure}

 We studied the hardness ratio distribution of the high-energy versus the medium-energy of \textit{Insight}-HXMT for all the bursts. The result is shown in Figure \ref{Fig8}, where the solid black curve is for the hardness ratio, and the dotted light blue curve is for SGR J1935+2154 X-ray pulse profile conducted from NuSTAR May 2 data. The hardness ratio curve has a significant double-peak structure, where the valley position is coincide with the X-ray pulse profile minimum point, while the peak position has an offset compared to the two peaks.
 
\section{Discussion}
\label{sect:discussion}
According to the above analysis results, we did not find an precise spin period from \textit{Insight}-HXMT observation data due to the bad Signal To Noise Ratio. The start time distribution of bursts in phase shows a uniform property. The energy spectrum of SGR J1935+2154 observed by Fermi/GBM and \textit{Insight}-HXMT at its intense activity stage after outbreak is complex, mainly double black-body spectrum and power law spectrum. In the phase space, the burst photons mainly come from the region corresponding to the two peaks of the persistent X-ray pulse profile. During the decay stage of the magnetar, corresponding to bursts detected by \textit{Insight}-HXMT, the burst photon distribution with the same spectrum type has obvious shift in phase relative to the intense stage of the outburst.

\begin{figure}
   \centering
   \includegraphics[width=8cm, angle=0]{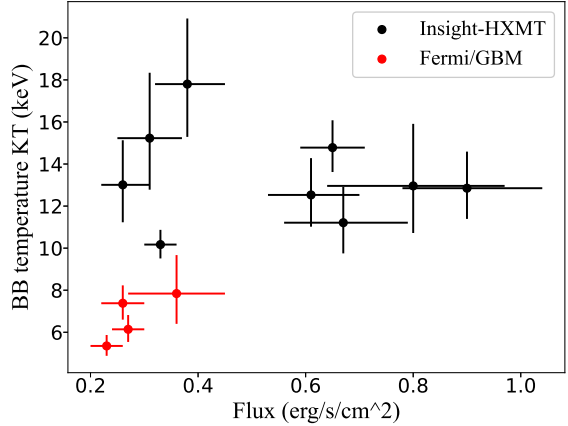}
   \caption{The relationship between temperature and flux of black body spectrum bursts, red dots represent Fermi/GBM bursts, all of which are during the activity period on April 27, and black dots are \textit{Insight}-HXMT data, all of which are after April 28.}
   \label{Fig9}
   \end{figure}

We analyzed the variation characteristics of burst temperature with flux for Fermi/GBM and \textit{Insight}-HXMT single black-body spectrum type, and the results are shown in Figure \ref{Fig9}, where the black dots are \textit{Insight}-HXMT BB bursts, and the red dots are Fermi/GBM bursts. It can be seen that the temperature of both Fermi/GBM and \textit{Insight}-HXMT black-body spectra does not evolve with the flux dramatically. It can also be seen that the surface temperature for region producing single black-body spectrum burst at the beginning of the outburst is relatively low compared to which the outbreak enters the decreasing stage, which is consistent with the distribution trend of the burst photons shown in Figure \ref{Fig6}, where the burst photons concentrated near the valley of the pulse profile at the initial stage but second maximum at the decay stage.

Although we did not find any significant structural features in the phase distribution of the burst start times, we found a tendency that the burst start times may align with the maxima peak of the X-ray pulse profile under certain conditions, such as the Fermi/GBM observations at the beginning of the outburst (see Figure \ref{Fig3} sub panel b), and the bursts with power law spectrum at different outburst activity stages (see Figure \ref{Fig5}), and also the bursts with double BB spectrum (see Figure \ref{Fig7}). In addition to this tendency of the burst start times, the photon arrival times of each burst have a more obvious concentrated structure align with the maxima peak. To date, except for the confirmed alignment phenomenon during the outburst of XTE J1810-197 (\citealt{Woods+etal+2008}), the alignment trend found in other magnetar reports during flux enhancement has been found to be inconclusive after in-depth study (\citealt{Ersin+etal+2017}). According to (\citealt{Elenbaas+etal+2018}) simulations, burst phase dependence is often affected by a number of external factors, such as observer angle and the location of the radiation area, beam bunching, and also requires a sufficient number of bursts to ensure complete sampling.
When the burst aligns near the maximum of the X-ray pulse profile, if it is a thermal burst, it may be generated from the surface of the neutron star. However, neither the capture of plasma fireball nor the hot spot generated by the bombardment of charged particles on the surface can explain the cause of the burst. It is possible that the thermal spectrum burst is generated by other different mechanisms at similar locations. If the observed phase alignment trend for non-thermal spectrum burst is true and not caused by observational effects or gravitational refraction, then it seems likely that the presence of a plasma fireball can explain this phenomenon, due to the twisted magnetic field the self-induction electric field lift particles off the magnetar surface, accelerate them, and produce radiation. The rising particles are trapped in the magnetosphere, forming a corona of plasma that heats the star's surface and creates thermal radiation, which may coincide in phase. However, only our observational data now cannot derive a definite conclusion, and we need more high-quality data to carry out further research.

\normalem
\begin{acknowledgements}
This work was partially supported by International Partnership Program of Chinese Academy of Sciences (Grant No.113111KYSB20190020) and by the National Key R\&D Program of China (2021YFA0718500) from the Minister of Science and Technology of China (MOST). The authors thank supports from the National Natural Science Foundation of China under Grants U1938109, U1838201, U1838202, 12173103, U2038101, U1938103, 12133007, U1938201 and 11733009.

\end{acknowledgements}
  
\bibliographystyle{raa}
\bibliography{ms2022-0387}

\end{document}